\documentclass[sigconf,10pt,authorversion]{acmart}

\usepackage[english]{babel}
\usepackage{blindtext}

\acmYear{2025}\copyrightyear{2025}
\acmConference[HOTOS'25]{Workshop in Hot Topics in Operating Systems}{May 14--16, 2025}{Banff, AB, Canada}
\acmBooktitle{Workshop in Hot Topics in Operating Systems (HOTOS 25), May 14--16, 2025, Banff, AB, Canada}
\acmDOI{10.1145/3713082.3730371}
\acmISBN{979-8-4007-1475-7/25/05}

\begin{document}

\title[Federated Maps to Enable Spatial Applications]{Uniting the World by Dividing it: Federated Maps to Enable Spatial Applications}

% \include{ccs}

% \keywords{Federated maps, spatial computing, distributed systems}

%\titlenote{Produces the permission block, and copyright information}
%\subtitle{Extended Abstract}

\author{Sagar Bharadwaj}
% \authornote{Note}
\orcid{0009-0003-3782-9912}
\affiliation{%
  \institution{Carnegie Mellon University}
  % \streetaddress{5000 Forbes Avenue}
  % \city{Pittsburgh} 
  % \state{PA} 
  % \postcode{15213}
}
% \email{skalasib@andrew.cmu.edu}

\author{Anthony Rowe}
\orcid{0000-0003-2332-9450}
\affiliation{%
  \institution{Carnegie Mellon University}
}

\author{Srinivasan Seshan}
\orcid{0000-0002-9508-2066}
\affiliation{%
  \institution{Carnegie Mellon University}
}

\renewcommand{\shortauthors}{Bharadwaj et al.}

% System name
\newcommand{\systemname}{\emph{OpenFLAME}}

% Notes
\newcommand{\sagar}[1]{\textcolor{red}{sagar: #1}}
\newcommand{\srini}[1]{\textcolor{blue}{Srini: #1}}

\begin{abstract}

The emergence of the Spatial Web -- the Web where content is tied to real-world locations has the potential to improve and enable many applications such as augmented reality, navigation, robotics, and more. The Spatial Web is missing a key ingredient that is impeding its growth -- a spatial naming system to resolve real-world locations to \textit{names}. Today's spatial naming systems are digital maps such as Google and Apple maps. These maps and the location-based services provided on top of these maps are primarily controlled by a few large corporations and mostly cover outdoor public spaces. Emerging classes of applications, such as persistent world-scale augmented reality, require detailed maps of both outdoor and indoor spaces. Existing centralized mapping infrastructures are proving insufficient for such applications because of the scale of cartography efforts required and the privacy of indoor map data.

In this paper, we present a case for a federated spatial naming system, or in other words, a federated mapping infrastructure. This enables disparate parties to manage and serve their own maps of physical regions and unlocks scalability of map management, isolation and privacy of maps. Map-related services such as address-to-location mapping, location-based search, and routing needs re-architecting to work on federated maps. We discuss some essential services and practicalities of enabling these services.

\end{abstract}
\maketitle
\section{Introduction}
\label{sec:intro}

Many have predicted the future of the Web to be the integration of Web-like content with the real-world. This overlay of virtual content on top of the physical world, which we refer to as the Spatial Web, holds promise for dramatically changing many applications from navigation, to engineering, education, and more.
While the vision is promising, the reality is that Spatial Web applications are difficult to create.

Looking at existing systems, we can see that a key enabler for such systems has often been the underlying infrastructure that makes it easy to discover and reference content. For the Web, the Domain Name System (DNS) provided a simple mechanism to convert human-readable names (domain names and URLs) to server IP addresses that can provide relevant content. The Spatial Web requires a similar infrastructure to relate human-readable names (e.g., The White House) with real-world locations and the content associated with that location (e.g., White House web page). In other words, we need a spatial naming system. Since this naming system translates names to physical locations, we use the term \textit{map} or \textit{mapping service} to refer to this naming system. 

The design of a naming system significantly influences and constrains the behavior of any distributed system built on top of it. For example, a naming system's mechanisms to add new entities to a distributed system can create bottlenecks to maintaining and scaling a system. The Web and the Internet at large were able to rapidly scale and incorporate a large number of hosts in their early days primarily due to the federated and pseudo-decentralized nature of the DNS. The federated design of the DNS allowed organizations to independently manage and control their level of participation on the Internet. The link between spatial naming and application constraints is no different and we can see this relationship in current deployed systems. 

%Peer-to-peer designs extend this further by relying on content lookup techniques that remove some of the structure and delegation requirements of DNS-like designs. 

Today's spatial naming systems are digital maps like Google and Apple maps. These digital maps are supported by centralized infrastructures and maintained by large corporations. Only the information that is gathered and exposed by organizations maintaining these centralized maps is available to applications, thereby severely limiting their functionality. In this paper, we make the case that to enable the rapid growth of the Spatial Web, we need an underlying spatial naming system or a map that is federated and can be independently controlled and maintained by disparate organizations.

% A naming system resolves a \textit{name} to the address of an entity that it refers to and is ubiquitously used in distributed systems. Names are strings of bits that refer to an \textit{entity} in a distributed system~\cite{}. For example, distributed databases use a centralized coordinator as the naming system to resolve row names to their location on a machine~\cite{chang2008bigtable, corbett2013spanner}. Peer-to-peer storage systems use a distributed hash table as their naming system to resolve file names to hosts~\cite{stoica2001chord}.  The Internet and the Web are enabled by the Domain Name System (DNS) that resolves human-readable names to host IP addresses~\cite{rfc1034}.

Extending spatial applications indoors is a use case that especially highlights the importance of a federated mapping infrastructure. Indoor maps contain sensitive information that needs to be owned and controlled by the owner of the physical space. Many organizations, such as stores, would benefit from providing accurate map data for applications such as product search~\cite{aukiLabsRetail, mapsPeopleMalls}, but would not be willing to publicly host detailed maps. Furthermore, the storage and cartography effort required for indoor spaces far outweighs that of outdoor maps. Some estimate that there could be more than 100 billion square feet (and growing) of indoor space in the world~\cite{commercialBuildFactsheet} and surveying this space will likely be impractical for any single centralized organization.

In this paper, we introduce \systemname{}\footnote{OpenFLAME stands for Open Federated Localization and Mapping Engine}, an architecture for a federated mapping infrastructure. \systemname{} is organized into `map servers' -- independent services deployed by potentially disparate parties that provide map data and location-based services confined to a physical region. \systemname{} provides the means to discover and tie these services together thereby providing a unified spatial naming system that can support spatial applications. A federated mapping infrastructure presents several challenges.

\begin{itemize}
    \item The maps are heterogeneous and can vary from one another in multiple ways. They can have different fidelities (eg. 2D and 3D maps), laid out in different coordinate systems (eg. indoor maps may not have exact latitude and longitude coordinates as its difficult to align them~\cite{felts2015location, nistIndoorLocalization}), and have different labels for common overlapping areas.
    \item Providing location-based services such as address-to-location lookup, routing, and search on top of federated maps requires a re-design of their architecture.
    \item We need a system that discovers map providers in a region. The discovery system should account for fuzziness of map boundaries and multiple ownership of physical regions.%In DNS, we boot-strap the discovery of the relevant name servers through the use of a well-known root server, the use of name-space delegation, and DNS glue records. However, these mechanisms do not translate well to spatial naming since we expect multiple map services (e.g., both Apple and Google) to overlap in their coverage, breaking DNS-like delegation mechanisms. 
\end{itemize}

In \S~\ref{sec:exampleApplication}, we give an example of a typical Spatial Web application that would be made possible by a federated mapping infrastructure. \S~\ref{sec:mapAbstraction} describes the abstraction of map and map servers. \S~\ref{sec:locServices} briefly describes the abstraction of location-based services typically used by spatial applications that would act as specifications for services to be enabled by \systemname{}. We also describe how the existing centralized model provides these services. \S~\ref{sec:practicalConsiderations} discusses some practical considerations for realizing the federated mapping infrastructure in practice.

%The foundational layer underlying all location-based services in \systemname{} is map server discovery. Client devices should be able to discover map servers that provide map data and location-based services for a region. Therefore, \systemname{} needs a mapping from regions to map server addresses. While a centralized spatial database would serve that purpose, the maintenance of this database reintroduces the same problems of scale raised earlier. As a result, we need a federated and scalable solution that is also easy to adopt. Map server discovery phase involves exclusively read queries, and does not require a full-fledged database with transaction processing, concurrency control, etc. We take advantage of the simplicity of our spatial queries and repurpose the Domain Name System (DNS) to work as our spatial database. Leveraging DNS for our purpose gives us access to its readily available ubiquitous caching mechanisms, thereby speeding up the discovery phase. The device's coarse location information is converted to a list of DNS-compatible domain names called geo-domains. We query the DNS with these geo-domains to get a list of map servers. Our system uses the existing DNS infrastructure without changes in implementation making \systemname{} easily adoptable.

\section{Example application}
\label{sec:exampleApplication}

To better understand the needs of future Spatial Web applications, we start by describing in detail what we consider a typical such application -- grocery store navigation. 

Let us consider a scenario where a user wishes to search for a product of interest, e.g., a particular flavor of seaweed, near their location. The application then provides the user with pedestrian navigation guidance to the exact shelf in a grocery store nearby that stocks the seaweed.

First, consider how an existing navigation application would support this task. The application would have to rely on a centralized database of destination locations and navigable paths exposed by a map provider such as Google. However, these databases are typically limited to street addresses and public landmarks. The seaweed in the store or the store aisles, for example, would not be a part of the map database unless the store has requested Google to maintain a database of the store inventory indexed by shelf locations (which would involve complicated and expensive integration between Google and the store's systems). Second, the navigation application would rely on a combination of technologies to determine the location of a user with respect to the map data, including GPS, image data from Google Street View, and WiFi/cellular signal strength. The availability of these technologies is limited to outdoor locations for GPS and to StreetView covered regions for image localization (typically public roads). Therefore, the navigation application would not work well within the store even if the Google Maps database included the user's product of interest.

Ideally, we would like the application to provide precise visual guidance along all steps of the path. Existing applications fail to meet this objective in multiple ways -- failing to provide precise guidance when localization is inaccurate and, in this case, failing to provide complete guidance as the requested seaweed is not part of Google Map's database. %The application failure is primarily due to the underlying map database and localization systems.

We envision \systemname{} would enable such spatial applications and more. In this case, the grocery store would maintain its own map independent of Google Maps. The grocery store's map would have a precise map of the store, along with its current inventory laid out against the local map. When the navigation application searches for the nearest seaweed, \systemname{} would discover the grocery store's map nearby and request from it the necessary shelf location information. Once the map and the location of interest within the map have been identified, the application can now estimate the navigation route from the user's current location on the street to the grocery store shelf. The route would be a combination of routes calculated by both Google Maps and the grocery store's map. Google Map route would lead the user to the storefront, while the grocery store's map would lead the user further to the shelf with the desired product. When outdoors, the application could rely on GPS or Street-view imagery to localize the user, but once indoors, the application would switch to the localization service provided by the grocery store's map to localize the user precisely within the store.

\subsection{Challenges}

The application described above exposes some challenges associated with realizing a federated mapping infrastructure. 

\textbf{Location-based services}: Notice that multiple location-based services were used by the application. \textit{Location-based search} to search for a product. \textit{Routing} to calculate the path on which to navigate the user. Finally, \textit{localization} to estimate where the user is with respect to the map while navigating. In a centralized infrastructure, the client application simply requests these services from a single entity that has access to all of the map data. In \systemname{}, the client device first has to discover relevant map servers and request the required services from these map servers, stitching the results if required. \S~\ref{sec:locServices} defines the main classes of location-based services. \S~\ref{subsec:federatedLocServices} shows how location-based services can be provided on top of a federated mapping infrastructure.

\textbf{Heterogeneity of maps}: Maps can be heterogeneous in multiple ways. Google Maps and the grocery store map in the above example can differ with respect to: \textit{Fidelity} -- the grocery store map can be more detailed with precise 3D information of shelves and what is stocked in them. Google Maps would be sparser in comparison. \textit{Coordinate frame of reference} -- the grocery store map might not be properly oriented in the geographic coordinate system of latitudes and longitudes as well as Google Maps. Aligning an indoor map accurately with the geographic coordinate system is a notoriously difficult problem~\cite{felts2015location, nistIndoorLocalization} and needs expensive survey equipment~\cite{rtkGnss, totalStation}. The data within the grocery store map is only precisely aligned against its own separate coordinate system. \textit{Common area labels} -- even if the maps have an overlapping region (e.g., some portion of the storefront), they might be labeled differently making alignment harder. Heterogeneity makes it challenging for map providers to jointly provide unified location-based services to applications.

\textbf{Overlapping maps}: In the above example, Google Maps also has sparse map data coverage of the region containing the grocery store. Therefore, multiple maps can overlap with each other. Note that in traditional naming systems such as DNS, there is no ambiguity concerning the ownership of a namespace. For example, the domain \texttt{www.google.com} is owned by Google and no other organization. However, the possibility of overlapping maps makes building a federated spatial naming system challenging.

\section{Map and Map servers}
\label{sec:mapAbstraction}

A map is a representation encoding relationships and attributes of spatial entities in a geographic region. While traditionally, a map refers to the visual representation of geographic features, in our context it is the data that underlies such visual representations. We adopt the widely used OpenStreetMap's data model for a map~\cite{osmElements}. A map has three major elements -- nodes, ways and relations. A \textit{node} represents a point on the map, defined by its coordinates within the map. A \textit{way} is an ordered list of nodes that defines a polyline and used to represent navigable paths, borders, rivers, etc. A \textit{relation} is used to represent a collection of related map elements -- nodes, ways or other relations. Each map element can have any metadata associated with it. 

A map server is a system that stores the map of a region and provides services such as search and routing on the map. The usefulness of a map server is determined by the services it implements. It can also impose fine-grained security and privacy policies on users and applications (\S~\ref{subsec:securityAndPrivacy}).

A map in \systemname{} is conceptually equivalent to a \textit{zone} in a traditional naming system like the DNS~\cite{rfc1035}. A \textit{DNS zone} is a portion of the DNS namespace that is independently managed by an organization. Similarly, a map is a portion of the spatial namespace that is independently managed by an organization. A map server is akin to a \textit{name server} in DNS parlance. However, there are two key differences between the concept of a DNS zone and a map as a zone. First, unlike DNS zones, the `boundary' of a map is fuzzy. Whenever a DNS zone, such as \texttt{google.com} is assigned to an organization, it is clear that even slight variations such as \texttt{googli.com} is outside the purview of the organization. However, in the spatial world, the polygonal boundary that might define the confines of a map is never exact. For example, the boundary of the grocery store map in the example in \S~\ref{sec:exampleApplication} might spill over to other stores nearby. This is inevitable especially for indoor maps, where finding out the exact geographic coordinates of walls and boundaries involves expensive surveying. Second, multiple maps may cover the same physical region. For example, \systemname{} should include both Google and Apple maps in its infrastructure. In DNS, ownership over a namespace is given exclusively to one zone. However, in the spatial naming system, multiple zones may cover the same physical region. In \S~\ref{subsec:mapServerDiscovery}, we briefly discuss how these challenges can be tackled in practice.
\section{Location-based services}
\label{sec:locServices}

%Most location-based applications rely on a set of common location-based services from map providers. In Section~\ref{subsec:locBasedServices}, we describe basic location-based services and how they are sufficient for most applications today. Section~\ref{subsec:centralizedMapModel} shows how existing centralized map infrastructures provide these services. 

% \subsection{Services}
% \label{subsec:locBasedServices}

A spatial naming system is only useful to spatial applications if we can provide location-based services on top of it. In this section, we will discuss existing location-based services that spatial applications use, which will act as specifications for the services that our naming system should enable.  In \S~\ref{subsec:federatedLocServices}, we discuss how these services can be implemented in practice on a federated map. 

Map providers offer a host of location-based services on their platform~\cite{googleMapsPlatform, mapboxDocumentation, osmServices, nianticLocate}. However, all of these services can be derived from a smaller set of base services. Some base services are different kinds of queries on the map data: simple address-to-location lookups, location-based search, and routing. Other base services are application enablers such as localization and map visualization. We describe these base services and how other services are derived from them.

\textbf{Forward and reverse geocode}: The process of converting a text-based address to a location on the map~\cite{Goldberg2007FromTT} is forward geocode. This underlies the sub-service of placing markers on the map given an address and is also a preliminary step in routing, as applications usually request a path between two addresses rather than two map nodes. The service that coverts a geographic location to a map node is called reverse geocode. It is the underlying service that supports click interactions on the map and snapping raw GPS coordinates to roads on the map while navigating ~\cite{googleRoadsAPI, mapboxMapMatchingAPI}.

\textbf{Location-based search}: Searching for map nodes using their metadata or features as keywords in or around a region is called location-based search. This service serves requests of the form "restaurants around me", "parking spot near the theater", etc. Map providers index map node features and metadata against their location to provide this service.

\textbf{Routing}: Routing is the service that provides a path from one map node to another. The path usually optimizes a metric such as distance, travel time, number of turns, toll price etc. Today's map providers typically run versions of shortest-path graph algorithms against their centralized map data to estimate optimized paths~\cite{bast2016route}. %In \systemname{}, optimal routing paths need to be jointly estimated by both the map providers and the client. The final path might involve passing through maps provided by multiple independent providers. 

\textbf{Localization}: The service that informs a device of its location and orientation with respect to a map is called localization. Today, since most map data aligns with a global geographic system, location-based applications typically rely on large-scale positioning systems like GPS, WiFi access points, cell towers, and Bluetooth~\cite{webGeolocation, appleGeolocation, androidGeolocation}.

%In a federated mapping system, we cannot expect all map providers to conform to the global geographic coordinate system so we allow maps that are in their own coordinate frame of reference. As a result, map providers have to provide their own localization system to enable devices to estimate their location with respect to the map. We describe how independent map providers can provide their own localization solutions to devices in Section~\ref{}.

%Numerous indoor localization systems have been established that use many technologies such as image-based localization~\cite{}, Ultra-wide band beacons~\cite{}, bluetooth beacons~\cite{}, etc. A map provider can a technology that is most suitable for their specific map. \systemname{} is agnostic to the underlying localization technology used by multiple map providers.

\textbf{Tile rendering}: Tile rendering powers interactive maps by delivering map tiles---2D images or 3D meshes---based on the user's latitude, longitude, and zoom level. As users drag or zoom, the tile server dynamically loads the appropriate tiles to update the view. %To create interactive map display service on top of \systemname{}, we need to discover maps hosted for a region of interest and the client has to stitch these different map coordinate systems together.

\subsection{Centralized map model}
\label{subsec:centralizedMapModel}

\begin{figure}
    \centering
    \includegraphics[width=0.7\linewidth]{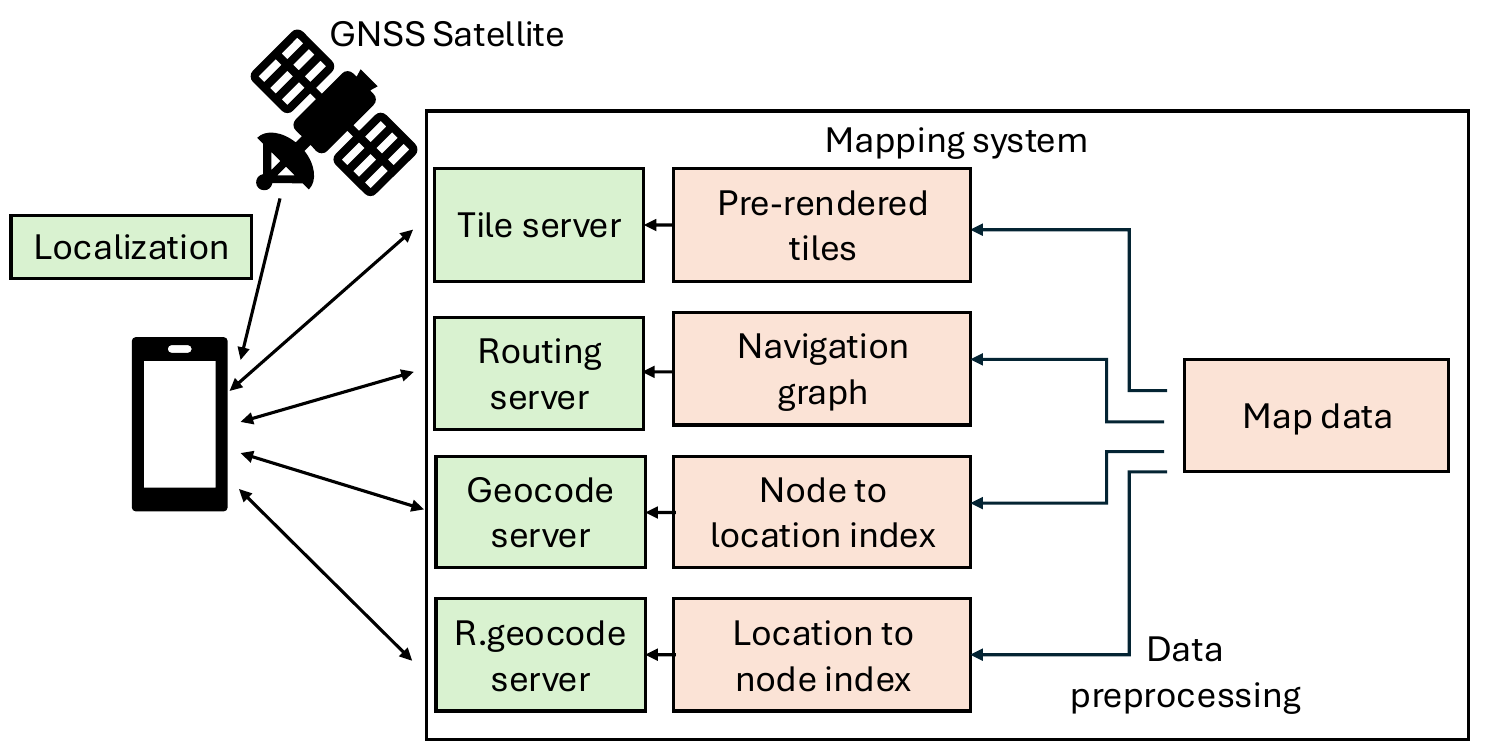}
    \caption{Centralized Architecture}
    \label{fig:centralizedArchitecture}
\end{figure}

Figure~\ref{fig:centralizedArchitecture} shows how today's centralized map infrastructures provide location-based services. The map data of the world is preprocessed into different forms required for each location-based service. For example, to provide the routing service, map data might be converted to a graph and then preprocessed using the contraction hierarchies algorithm which makes routing queries faster to compute~\cite{contractionHierarchies}. The tile rendering service might pre-render tiles corresponding to latitudes, longitudes and zoom levels even before they are requested by any client. Geocode, reverse geocode, and location-based search would involve indexing map nodes and their metadata against geographic coordinates. 

The API calls to each service would then use the pre-processed data to serve requests. For example, the tile service API would fetch appropriate tiles from the pre-rendered set and serve them to the client.
\section{Practical Considerations}
\label{sec:practicalConsiderations}

\begin{figure}
    \centering
    \includegraphics[width=0.7\linewidth]{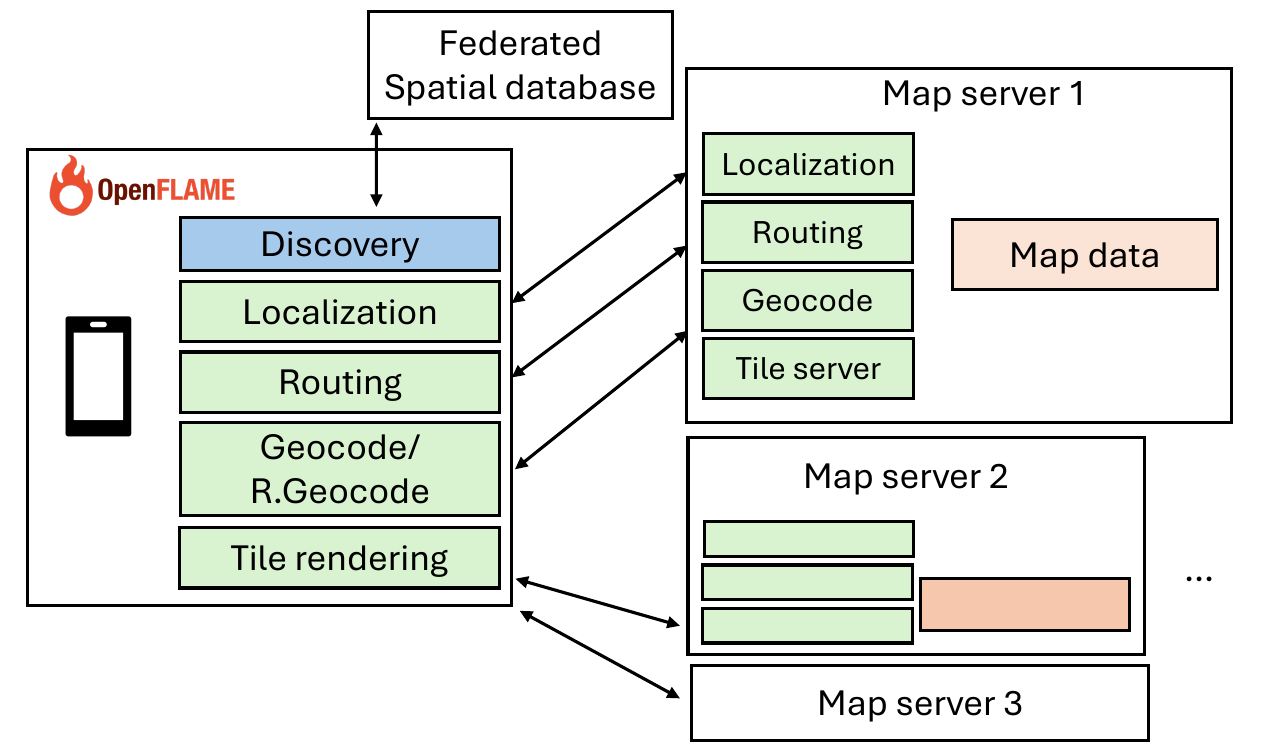}
    \caption{\systemname{} Architecture}
    \label{fig:ourArchitecture}
\end{figure}

Figure~\ref{fig:ourArchitecture} shows the \systemname{} architecture. \systemname{} client relies on a discovery mechanism to identify all the map providers in a region. \S~\ref{subsec:mapServerDiscovery} discusses how the challenges presented in \S~\ref{sec:mapAbstraction} can be tackled in practice while implementing a map server discovery system. \S~\ref{subsec:federatedLocServices} discusses the split of responsibilities between the client and map servers for providing services described in \S~\ref{sec:locServices}. Federation enables a finer grained security and privacy model discussed in \S~\ref{subsec:securityAndPrivacy}.

%Figure~\ref{fig:ourArchitecture} shows the \systemname{} architecture. The implementation of location-based services is split between the client and the map servers (Section~\ref{subsec:federatedLocServices}). The client also implements a discovery mechanism to discover map servers relevant to a region to be able to request location-based services from them (Section~\ref{subsec:mapServerDiscovery}). Federating mapping services enables a finer grain security and privacy model which we describe in Section~\ref{subsec:securityAndPrivacy}.

\subsection{Map server discovery}
\label{subsec:mapServerDiscovery}

The foundation that underlies all location-based services is the discovery layer. The discovery system would essentially maintain the data mapping locations to map servers in that location. The discovery query would involve the coarse location of the device obtained from ubiquitous sources like the GPS. The discovery system would then respond to the query with a list of map providers for the region. %To ensure that \systemname{} remains federated, the underlying discovery system should also be federated. 

% \srini{I don't think this is the right description -- describe the query you want and the result you want. Also describe the information that is reasonable for each component to maintain. Draw parallels to things like resolvers need to know the local name server and local name servers must know DNS root}

% \sagar{1. Define the concept of zones. Draw parallels to the DNS zone concept. 2. Describe differences with the DNS zone concept. 3. One way zones is to have precise polygons, but we don't want that -- uncertainty because of partial knowledge. 4. Describe zone Hierarchy as its better. }

%An approach to implementing the discovery system would be a federated spatial database. The database would be split into zones defined by polygonal boundaries. A \textit{root table} would contain pointers to all \textit{zone tables}. Each zone table would contain records pointing to map servers providing services within their region. The database records in each zone table would be independently maintained by an organization. This structure is very similar to today's DNS. For example, a university would maintain a zone table for the region spanning the university premises. The zone table would contain records pointing to map servers maintained by individual departments for their own buildings.

Several existing systems such as spatial databases~\cite{geo_fire, PostGIS, mongodb}, Geographic Information Systems (GIS)~\cite{arc_gis, carto}, and the Intentional Naming System (INS)~\cite{adjie1999design} can be used for spatial discovery. %The discovery system can be structured as a federated database with a \textit{root table} pointing to \textit{zone tables}---discovery systems maintained by independent organizations for their own physical zones, similar to today's DNS.
However, deploying the infrastructure required for these systems from the ground up and maintaining them would be a major impediment to the adoption of the Spatial Web. Let us consider the following observations about the discovery system that allows us to use an existing widely deployed infrastructure---the DNS. 1. The fuzziness of map boundaries discussed in \S~\ref{sec:mapAbstraction} does not require a database that maintains precise polygonal boundaries. 2. Discovery queries are exclusively read queries and do not require a full-fledged database with transaction processing. 3. The address of the map servers are not expected to change frequently so the system would benefit from a ubiquitous caching mechanism. 

To repurpose the already federated DNS to work as the spatial database, we can leverage spatial indexing systems (e.g., S2~\cite{s2home}, H3~\cite{h3}) to convert locations to hierarchical domain names. A polygonal region, or a zone, can be approximated by a collection of domain names. Coarse location in the form of latitude and longitude can also be converted to a domain name. Therefore, the discovery query would be a simple domain look-up on the DNS. Leveraging DNS for our purpose gives us access to its ubiquitous caching mechanisms, large-scale deployments, and infrastructure. Previously, the Spatial Name System~\cite{spatialNameSystem} has also considered using the DNS to assign and resolve hierarchical location-based names. However, their use of civic addresses as domain names while maintaining geodetic location as part of record data results in inefficient location-based discovery process.

\subsection{Location-based services}
\label{subsec:federatedLocServices}

In this section, we discuss split of responsibilities between \systemname{} client and map servers in providing location-based services from \S~\ref{sec:locServices} on top of a federated map. %A map server can choose to implement and expose any of these services and none of them are mandatory. The services provided by a map server limits how its map data can be used. %For example, if a map server only exposes a tile service without providing localization, its map data can be rendered on an interactive map but a user cannot navigate through it when physically present in that space. Below we describe how \systemname{} client and map servers can work jointly to provide location-based services to spatial applications.

\textbf{Geocode}: Given a text string of a hierarchical address, the client first uses the geocode service of a large world-map provider (e.g., OpenStreetMap~\cite{openStreetMap}) to get the coarse location of a part of the address. The client then discovers finer map servers in the coarse location which search in their own maps for the exact address. %For example, let us consider the address ``Printer, Gandalf's office, White House, DC, USA''. While OpenStreetMap's data might not contain the location of the printer in Gandalf's office, it will give the location of the White House. \systemname{} client can then discover the White House map server and search for the location of the printer within that map server.

\textbf{Reverse geocode and location-based search}: Searching for map nodes around a location would begin by the client discovering map servers around a given location. The client would then ask each map server to search for the relevant items within their maps and return relevant results, if any. The client would then rank results from multiple map servers and present them to the application. %Reverse geocode would just present the closest map node from the closest map server.

\textbf{Routing}: The client first obtains the location of the source and destination addresses using the Geocode service described above. Then it discovers all the map servers that lie along the way from the source to the destination. Each map server would calculate the route that is relevant for the region that they cover. The client would collect paths from all relevant map servers, and stitch them together such that the final path optimizes a metric of interest.

\textbf{Localization}:
The process of localizing a device starts with the \systemname{} client discovering map servers in the location. The client might discover multiple overlapping servers or even unrelated maps because of the coarseness of the discovery process. Once the map servers are discovered, the client sends them `location cues' collected by the device sensors -- images, beacon signals, fiduciary tag scans, etc. The location cue sent to the map server depends on the localization technology advertised by the server. %For example, if the server advertises image-based localization, it would accept images. 
The map servers accept location cues, localize the device within their map, and return the results to the client. The client then selects the best one by comparing these results with its own IMU (Inertial Measurement Unit) sensors or local SLAM (Simultaneous Localization and Mapping) algorithm~\cite{SLAMProbabilisticRobotics}. The most plausible result is returned to the application.

\textbf{Tile rendering}: Each map server would expose a visual representation of its map data as 2D images, 3D meshes or other forms. The client would download these representations from multiple discovered map servers and stitch them together before showing them to the user. For example, stitching together map data in different coordinates and projection systems can be done using manual correspondences between maps (e.g., MapCruncher~\cite{elson2007mapcruncher}).

% \subsection{Spatial Application}
% \label{subsec:application}

\subsection{Security and privacy model}
\label{subsec:securityAndPrivacy}

Unlike in the case of a centralized map, map providers in \systemname{} can control access to their data and services in fine-grained ways as they can implement separate authentication processes for each of the services and map data. \textit{User-level control} -- A map server covering a university, for example, may only serve users who can authenticate with the university's email address. This ensures users who are not from the university cannot get fine-grained map data. \textit{Service-level control} -- A map server, for example, may provide its tile service to a large set of users so they can view the map. However, it may choose to provide localization service only to a small set of users who are supposed to have physical access to the place. \textit{Application-level control} -- A university might provide localization service only if it comes from the campus navigation application and trust that the application has implemented its own way of authenticating users.
\section{Conclusion}

In this paper we present the need for a federated spatial naming system, or a map, to enable emerging Spatial Web applications. We then discuss how location-based services that are primarily utilized by today's spatial applications can be provided on top of a federated map. We hope that this paper encourages the systems community to think about the underlying infrastructure required to enable the next generation of the Web.

\bibliographystyle{ACM-Reference-Format}
\bibliography{reference}


\begin{thebibliography}{33}

%%% the Web address.

\ifx \showCODEN    \undefined \def \showCODEN     #1{\unskip}     \fi
\ifx \showDOI      \undefined \def \showDOI       #1{#1}\fi
\ifx \showISBNx    \undefined \def \showISBNx     #1{\unskip}     \fi
\ifx \showISBNxiii \undefined \def \showISBNxiii  #1{\unskip}     \fi
\ifx \showISSN     \undefined \def \showISSN      #1{\unskip}     \fi
\ifx \showLCCN     \undefined \def \showLCCN      #1{\unskip}     \fi
\ifx \shownote     \undefined \def \shownote      #1{#1}          \fi
\ifx \showarticletitle \undefined \def \showarticletitle #1{#1}   \fi
\ifx \showURL      \undefined \def \showURL       {\relax}        \fi
% invisible to TeX
\providecommand\bibfield[2]{#2}
\providecommand\bibinfo[2]{#2}
\providecommand\natexlab[1]{#1}
\providecommand\showeprint[2][]{arXiv:#2}

\bibitem[\protect\citeauthoryear{Adjie-Winoto, Schwartz, Balakrishnan, and Lilley}{Adjie-Winoto et~al\mbox{.}}{1999}]%
        {adjie1999design}
\bibfield{author}{\bibinfo{person}{William Adjie-Winoto}, \bibinfo{person}{Elliot Schwartz}, \bibinfo{person}{Hari Balakrishnan}, {and} \bibinfo{person}{Jeremy Lilley}.} \bibinfo{year}{1999}\natexlab{}.
\newblock \showarticletitle{The design and implementation of an intentional naming system}. In \bibinfo{booktitle}{{\em Proceedings of the Seventeenth ACM Symposium on Operating Systems Principles}} {\em (\bibinfo{series}{SOSP '99})}. \bibinfo{publisher}{Association for Computing Machinery}, \bibinfo{address}{New York, NY, USA}, \bibinfo{pages}{186–201}.
\newblock
\showISBNx{1581131402}
\showDOI{%
\url{https://doi.org/10.1145/319151.319164}}


\bibitem[\protect\citeauthoryear{Apple}{Apple}{2024}]%
        {appleGeolocation}
\bibfield{author}{\bibinfo{person}{Apple}.} \bibinfo{year}{2024}\natexlab{}.
\newblock \bibinfo{title}{{Core Location Apple}}.
\newblock \bibinfo{howpublished}{\url{https://developer.apple.com/documentation/corelocation/}}.   (\bibinfo{year}{2024}).
\newblock
\newblock
\shownote{Accessed: 2024-09-01.}


\bibitem[\protect\citeauthoryear{AukiLabs}{AukiLabs}{2024}]%
        {aukiLabsRetail}
\bibfield{author}{\bibinfo{person}{AukiLabs}.} \bibinfo{year}{2024}\natexlab{}.
\newblock \bibinfo{title}{{Auki labs retail solutions}}.
\newblock \bibinfo{howpublished}{\url{https://www.aukilabs.com/solutions/industries/retail}}.   (\bibinfo{year}{2024}).
\newblock
\newblock
\shownote{Accessed: 2024-09-17.}


\bibitem[\protect\citeauthoryear{Bast, Delling, Goldberg, M{\"u}ller-Hannemann, Pajor, Sanders, Wagner, and Werneck}{Bast et~al\mbox{.}}{2016}]%
        {bast2016route}
\bibfield{author}{\bibinfo{person}{Hannah Bast}, \bibinfo{person}{Daniel Delling}, \bibinfo{person}{Andrew Goldberg}, \bibinfo{person}{Matthias M{\"u}ller-Hannemann}, \bibinfo{person}{Thomas Pajor}, \bibinfo{person}{Peter Sanders}, \bibinfo{person}{Dorothea Wagner}, {and} \bibinfo{person}{Renato~F. Werneck}.} \bibinfo{year}{2016}\natexlab{}.
\newblock \bibinfo{booktitle}{{\em Route Planning in Transportation Networks}}.
\newblock \bibinfo{publisher}{Springer International Publishing}, \bibinfo{address}{Cham}, \bibinfo{pages}{19--80}.
\newblock
\showISBNx{978-3-319-49487-6}
\showDOI{%
\url{https://doi.org/10.1007/978-3-319-49487-6_2}}


\bibitem[\protect\citeauthoryear{CARTO}{CARTO}{2025}]%
        {carto}
\bibfield{author}{\bibinfo{person}{CARTO}.} \bibinfo{year}{2025}\natexlab{}.
\newblock \bibinfo{title}{CARTO}.
\newblock \bibinfo{howpublished}{\url{https://carto.com/}}.   (\bibinfo{year}{2025}).
\newblock
\newblock
\shownote{Online. Accessed: April 2025.}


\bibitem[\protect\citeauthoryear{Center~for Sustainable~Systems}{Center~for Sustainable~Systems}{2024}]%
        {commercialBuildFactsheet}
\bibfield{author}{\bibinfo{person}{University of~Michigan Center~for Sustainable~Systems}.} \bibinfo{year}{2024}\natexlab{}.
\newblock \bibinfo{title}{{Commercial Buildings Factsheet}}.
\newblock \bibinfo{howpublished}{\url{https://css.umich.edu/publications/factsheets/built-environment/commercial-buildings-factsheet}}.   (\bibinfo{year}{2024}).
\newblock
\newblock
\shownote{Accessed: 2024-09-01.}


\bibitem[\protect\citeauthoryear{Docs}{Docs}{2024}]%
        {webGeolocation}
\bibfield{author}{\bibinfo{person}{MDN~Web Docs}.} \bibinfo{year}{2024}\natexlab{}.
\newblock \bibinfo{title}{{Geolocation MDN}}.
\newblock \bibinfo{howpublished}{\url{https://developer.mozilla.org/en-US/docs/Web/API/Geolocation_API}}.   (\bibinfo{year}{2024}).
\newblock
\newblock
\shownote{Accessed: 2024-09-01.}


\bibitem[\protect\citeauthoryear{Elson, Howell, and Douceur}{Elson et~al\mbox{.}}{2007}]%
        {elson2007mapcruncher}
\bibfield{author}{\bibinfo{person}{Jeremy Elson}, \bibinfo{person}{Jon Howell}, {and} \bibinfo{person}{John~R Douceur}.} \bibinfo{year}{2007}\natexlab{}.
\newblock \showarticletitle{MapCruncher: integrating the world's geographic information}.
\newblock \bibinfo{journal}{{\em ACM SIGOPS Operating Systems Review\/}} \bibinfo{volume}{41}, \bibinfo{number}{2} (\bibinfo{year}{2007}), \bibinfo{pages}{50--59}.
\newblock


\bibitem[\protect\citeauthoryear{Esri}{Esri}{2025}]%
        {arc_gis}
\bibfield{author}{\bibinfo{person}{Esri}.} \bibinfo{year}{2025}\natexlab{}.
\newblock \bibinfo{title}{ArcGIS}.
\newblock \bibinfo{howpublished}{\url{https://www.arcgis.com/index.html}}.   (\bibinfo{year}{2025}).
\newblock
\newblock
\shownote{Online. Accessed: April 2025.}


\bibitem[\protect\citeauthoryear{Felts, Leh, McElvaney, and Orr}{Felts et~al\mbox{.}}{2015}]%
        {felts2015location}
\bibfield{author}{\bibinfo{person}{Ryan Felts}, \bibinfo{person}{Marc Leh}, \bibinfo{person}{Tracy McElvaney}, {and} \bibinfo{person}{Dereck Orr}.} \bibinfo{year}{2015}\natexlab{}.
\newblock \bibinfo{booktitle}{{\em Location-Based Services R\&D Roadmap}}.
\newblock \bibinfo{publisher}{US Department of Commerce, National Institute of Standards and Technology}, \bibinfo{address}{US}.
\newblock


\bibitem[\protect\citeauthoryear{Geisberger, Sanders, Schultes, and Vetter}{Geisberger et~al\mbox{.}}{2012}]%
        {contractionHierarchies}
\bibfield{author}{\bibinfo{person}{Robert Geisberger}, \bibinfo{person}{Peter Sanders}, \bibinfo{person}{Dominik Schultes}, {and} \bibinfo{person}{Christian Vetter}.} \bibinfo{year}{2012}\natexlab{}.
\newblock \showarticletitle{Exact Routing in Large Road Networks Using Contraction Hierarchies}.
\newblock \bibinfo{journal}{{\em Transportation Science\/}} \bibinfo{volume}{46}, \bibinfo{number}{3} (\bibinfo{date}{Aug.} \bibinfo{year}{2012}), \bibinfo{pages}{388–404}.
\newblock
\showISSN{1526-5447}
\showDOI{%
\url{https://doi.org/10.1287/trsc.1110.0401}}


\bibitem[\protect\citeauthoryear{Gibb, Madhavapeddy, and Crowcroft}{Gibb et~al\mbox{.}}{2023}]%
        {spatialNameSystem}
\bibfield{author}{\bibinfo{person}{Ryan Gibb}, \bibinfo{person}{Anil Madhavapeddy}, {and} \bibinfo{person}{Jon Crowcroft}.} \bibinfo{year}{2023}\natexlab{}.
\newblock \showarticletitle{Where on Earth is the Spatial Name System?}. In \bibinfo{booktitle}{{\em Proceedings of the 22nd ACM Workshop on Hot Topics in Networks}} {\em (\bibinfo{series}{HotNets '23})}. \bibinfo{publisher}{Association for Computing Machinery}, \bibinfo{address}{New York, NY, USA}, \bibinfo{pages}{79–86}.
\newblock
\showISBNx{9798400704154}
\showDOI{%
\url{https://doi.org/10.1145/3626111.3628210}}


\bibitem[\protect\citeauthoryear{Goldberg, Wilson, and Knoblock}{Goldberg et~al\mbox{.}}{2007}]%
        {Goldberg2007FromTT}
\bibfield{author}{\bibinfo{person}{Daniel~W. Goldberg}, \bibinfo{person}{John~P. Wilson}, {and} \bibinfo{person}{Craig~A. Knoblock}.} \bibinfo{year}{2007}\natexlab{}.
\newblock \showarticletitle{From Text to Geographic Coordinates: The Current State of Geocoding}.
\newblock \bibinfo{journal}{{\em Urisa Journal\/}}  \bibinfo{volume}{19} (\bibinfo{year}{2007}), \bibinfo{pages}{33}.
\newblock
\showURL{%
\url{https://api.semanticscholar.org/CorpusID:5517082}}


\bibitem[\protect\citeauthoryear{Google}{Google}{2024a}]%
        {googleMapsPlatform}
\bibfield{author}{\bibinfo{person}{Google}.} \bibinfo{year}{2024}\natexlab{a}.
\newblock \bibinfo{title}{{Google Maps Platform}}.
\newblock \bibinfo{howpublished}{\url{https://developers.google.com/maps}}.   (\bibinfo{year}{2024}).
\newblock
\newblock
\shownote{Accessed: 2025-01-14.}


\bibitem[\protect\citeauthoryear{Google}{Google}{2024b}]%
        {s2home}
\bibfield{author}{\bibinfo{person}{Google}.} \bibinfo{year}{2024}\natexlab{b}.
\newblock \bibinfo{title}{{S2 library}}.
\newblock \bibinfo{howpublished}{\url{http://s2geometry.io/}}.   (\bibinfo{year}{2024}).
\newblock
\newblock
\shownote{Accessed: 2024-09-01.}


\bibitem[\protect\citeauthoryear{Google}{Google}{2025a}]%
        {androidGeolocation}
\bibfield{author}{\bibinfo{person}{Google}.} \bibinfo{year}{2025}\natexlab{a}.
\newblock \bibinfo{title}{{Fused Location Provider Android}}.
\newblock \bibinfo{howpublished}{\url{https://developer.android.com/develop/sensors-and-location/location/retrieve-current}}.   (\bibinfo{year}{2025}).
\newblock
\newblock
\shownote{Accessed: 2024-09-01.}


\bibitem[\protect\citeauthoryear{Google}{Google}{2025b}]%
        {geo_fire}
\bibfield{author}{\bibinfo{person}{Google}.} \bibinfo{year}{2025}\natexlab{b}.
\newblock \bibinfo{title}{GeoFire for Javascript}.
\newblock \bibinfo{howpublished}{\url{https://github.com/firebase/geofire-js}}.   (\bibinfo{year}{2025}).
\newblock
\newblock
\shownote{Online. Accessed: April 2025.}


\bibitem[\protect\citeauthoryear{Inc.}{Inc.}{2025}]%
        {mongodb}
\bibfield{author}{\bibinfo{person}{MongoDB Inc.}} \bibinfo{year}{2025}\natexlab{}.
\newblock \bibinfo{title}{MongoDB}.
\newblock \bibinfo{howpublished}{\url{https://www.mongodb.com/}}.   (\bibinfo{year}{2025}).
\newblock
\newblock
\shownote{Online. Accessed: April 2025.}


\bibitem[\protect\citeauthoryear{Mapbox}{Mapbox}{2024a}]%
        {mapboxMapMatchingAPI}
\bibfield{author}{\bibinfo{person}{Mapbox}.} \bibinfo{year}{2024}\natexlab{a}.
\newblock \bibinfo{title}{{Map Mathcing API}}.
\newblock \bibinfo{howpublished}{\url{https://docs.mapbox.com/api/navigation/map-matching/}}.   (\bibinfo{year}{2024}).
\newblock
\newblock
\shownote{Accessed: 2025-01-14.}


\bibitem[\protect\citeauthoryear{Mapbox}{Mapbox}{2024b}]%
        {mapboxDocumentation}
\bibfield{author}{\bibinfo{person}{Mapbox}.} \bibinfo{year}{2024}\natexlab{b}.
\newblock \bibinfo{title}{{Mapbox Documentation}}.
\newblock \bibinfo{howpublished}{\url{https://docs.mapbox.com/}}.   (\bibinfo{year}{2024}).
\newblock
\newblock
\shownote{Accessed: 2025-01-14.}


\bibitem[\protect\citeauthoryear{maps platform}{maps platform}{2024}]%
        {googleRoadsAPI}
\bibfield{author}{\bibinfo{person}{Google maps platform}.} \bibinfo{year}{2024}\natexlab{}.
\newblock \bibinfo{title}{{Roads API}}.
\newblock \bibinfo{howpublished}{\url{https://developers.google.com/maps/documentation/roads}}.   (\bibinfo{year}{2024}).
\newblock
\newblock
\shownote{Accessed: 2025-01-14.}


\bibitem[\protect\citeauthoryear{MapsPeople}{MapsPeople}{2024}]%
        {mapsPeopleMalls}
\bibfield{author}{\bibinfo{person}{MapsPeople}.} \bibinfo{year}{2024}\natexlab{}.
\newblock \bibinfo{title}{{MapsIndoors for Malls}}.
\newblock \bibinfo{howpublished}{\url{https://www.mapspeople.com/industries/malls}}.   (\bibinfo{year}{2024}).
\newblock
\newblock
\shownote{Accessed: 2024-09-17.}


\bibitem[\protect\citeauthoryear{Mockapetris}{Mockapetris}{1987}]%
        {rfc1035}
\bibfield{author}{\bibinfo{person}{Paul~V Mockapetris}.} \bibinfo{year}{1987}\natexlab{}.
\newblock \bibinfo{title}{{RFC1034: DOMAIN names -- Implementation and Specification}}.
\newblock   (\bibinfo{year}{1987}).
\newblock


\bibitem[\protect\citeauthoryear{Niantic}{Niantic}{2024}]%
        {nianticLocate}
\bibfield{author}{\bibinfo{person}{Niantic}.} \bibinfo{year}{2024}\natexlab{}.
\newblock \bibinfo{title}{{Niantic Lightship -- Locate}}.
\newblock \bibinfo{howpublished}{\url{https://www.nianticspatial.com/locate}}.   (\bibinfo{year}{2024}).
\newblock
\newblock
\shownote{Accessed: 2025-01-14.}


\bibitem[\protect\citeauthoryear{NIST}{NIST}{2024}]%
        {nistIndoorLocalization}
\bibfield{author}{\bibinfo{person}{NIST}.} \bibinfo{year}{2024}\natexlab{}.
\newblock \bibinfo{title}{{Indoor Localization at NIST}}.
\newblock \bibinfo{howpublished}{\url{https://www.nist.gov/ctl/pscr/indoor-localization-nist}}.   (\bibinfo{year}{2024}).
\newblock
\newblock
\shownote{Accessed: 2024-09-17.}


\bibitem[\protect\citeauthoryear{OpenStreetMap}{OpenStreetMap}{2024a}]%
        {osmServices}
\bibfield{author}{\bibinfo{person}{OpenStreetMap}.} \bibinfo{year}{2024}\natexlab{a}.
\newblock \bibinfo{title}{{List of OSM-based services}}.
\newblock \bibinfo{howpublished}{\url{https://wiki.openstreetmap.org/wiki/List_of_OSM-based_services}}.   (\bibinfo{year}{2024}).
\newblock
\newblock
\shownote{Accessed: 2025-01-14.}


\bibitem[\protect\citeauthoryear{OpenStreetMap}{OpenStreetMap}{2024b}]%
        {osmElements}
\bibfield{author}{\bibinfo{person}{OpenStreetMap}.} \bibinfo{year}{2024}\natexlab{b}.
\newblock \bibinfo{title}{{OpenStreetMap Elements}}.
\newblock \bibinfo{howpublished}{\url{https://wiki.openstreetmap.org/wiki/Elements}}.   (\bibinfo{year}{2024}).
\newblock
\newblock
\shownote{Accessed: 2025-01-14.}


\bibitem[\protect\citeauthoryear{OpenStreetMap}{OpenStreetMap}{2025}]%
        {openStreetMap}
\bibfield{author}{\bibinfo{person}{OpenStreetMap}.} \bibinfo{year}{2025}\natexlab{}.
\newblock \bibinfo{title}{OpenStreetMap}.
\newblock \bibinfo{howpublished}{\url{https://www.openstreetmap.org}}.   (\bibinfo{year}{2025}).
\newblock
\newblock
\shownote{Online. Accessed: April 2025.}


\bibitem[\protect\citeauthoryear{PostGIS and OSGeo}{PostGIS and OSGeo}{2025}]%
        {PostGIS}
\bibfield{author}{\bibinfo{person}{PostGIS} {and} \bibinfo{person}{OSGeo}.} \bibinfo{year}{2025}\natexlab{}.
\newblock \bibinfo{title}{PostGIS}.
\newblock \bibinfo{howpublished}{\url{https://postgis.net/}}.   (\bibinfo{year}{2025}).
\newblock
\newblock
\shownote{Online. Accessed: April 2025.}


\bibitem[\protect\citeauthoryear{Thrun, Burgard, and Fox}{Thrun et~al\mbox{.}}{2006}]%
        {SLAMProbabilisticRobotics}
\bibfield{author}{\bibinfo{person}{Sebastian Thrun}, \bibinfo{person}{Woflram Burgard}, {and} \bibinfo{person}{Dieter Fox}.} \bibinfo{year}{2006}\natexlab{}.
\newblock \bibinfo{booktitle}{{\em Probabilistic Robotics}}.
\newblock \bibinfo{publisher}{MIT Press}, \bibinfo{address}{Cambridge, Masachussets, London, England}, Chapter~10, \bibinfo{pages}{309--334}.
\newblock


\bibitem[\protect\citeauthoryear{Uber}{Uber}{2024}]%
        {h3}
\bibfield{author}{\bibinfo{person}{Uber}.} \bibinfo{year}{2024}\natexlab{}.
\newblock \bibinfo{title}{{H3}}.
\newblock \bibinfo{howpublished}{\url{https://h3geo.org/}}.   (\bibinfo{year}{2024}).
\newblock
\newblock
\shownote{Accessed: 2024-09-01.}


\bibitem[\protect\citeauthoryear{Wikipedia}{Wikipedia}{2024a}]%
        {rtkGnss}
\bibfield{author}{\bibinfo{person}{Wikipedia}.} \bibinfo{year}{2024}\natexlab{a}.
\newblock \bibinfo{title}{{RTK GNSS}}.
\newblock \bibinfo{howpublished}{\url{https://en.wikipedia.org/wiki/Real-time_kinematic_positioning}}.   (\bibinfo{year}{2024}).
\newblock
\newblock
\shownote{Accessed: 2024-09-01.}


\bibitem[\protect\citeauthoryear{Wikipedia}{Wikipedia}{2024b}]%
        {totalStation}
\bibfield{author}{\bibinfo{person}{Wikipedia}.} \bibinfo{year}{2024}\natexlab{b}.
\newblock \bibinfo{title}{{Total Station}}.
\newblock \bibinfo{howpublished}{\url{https://en.wikipedia.org/wiki/Total_station}}.   (\bibinfo{year}{2024}).
\newblock
\newblock
\shownote{Accessed: 2024-09-01.}


\end{thebibliography}

\end{document}